\documentclass[reprint,amsmath,amssymb, aps]{revtex4-2}
\usepackage{multirow}
\usepackage{amssymb}
\usepackage{amsmath, graphicx}
\usepackage{dcolumn}
\usepackage{bm}
\usepackage{enumerate}
\usepackage{slashed}
\usepackage{epstopdf}
\usepackage[unicode]{hyperref}
\usepackage{csquotes}
\usepackage[usenames]{color}

\newcommand{\be}{\begin{equation}}
\newcommand{\ee}{\end{equation}}
\newcommand{\bea}{\begin{eqnarray}}
\newcommand{\eea}{\end{eqnarray}}
\newcommand{\ben}{\begin{enumerate}}
\newcommand{\een}{\end{enumerate}}
\newcommand{\bde}{\begin{widetext}}
\newcommand{\ede}{\end{widetext}}

\newcommand{\Tr}{\mathrm{Tr}}

\newcommand{\bc}{\begin{center}}
\newcommand{\ec}{\end{center}}

\usepackage{braket}

\begin{document}

\title{\boldmath Time-System Entanglement and Special Relativity}

\author{Ngo Phuc Duc Loc} \thanks{Email: locngo148@gmail.com}

\affiliation{Department of Physics and Astronomy, University of New Mexico, Albuquerque, New Mexico 87131, USA}

\begin{abstract}
We know that space and time are treated almost equally in classical physics, but we also know that this is not the case for quantum mechanics. A quantum description of both space and time is important to really understand the quantum nature of reality. The Page-Wootters mechanism of quantum time is a promising starting point, according to which the evolution of the quantum system is described by the entanglement between it and quantum temporal degrees of freedom. In this paper, we consider a qubit clock that is entangled with a quantum system due to the Wigner rotation induced by Lorentz transformation. We study how this time-system entanglement depends on the rapidity of the Lorentz boost. We consider the case of a spin-1/2 particle with Gaussian momentum distribution as a concrete example. We also compare the time-system entanglement entropy with the spin-momentum entanglement entropy and find that the former is smaller than the latter.
\end{abstract}
\maketitle

\section{Introduction}
One of the biggest goals of physicists today is to find a good quantum theory of gravity that can reconcile the weirdness of quantum mechanics and the deterministic features of gravity. There is a  traditional approach to quantize gravity \cite{joe} and a more recent radical approach to ``gravitize" quantum mechanics \cite{sean1,sean2,giddings}. None of these proposals is clear-cut yet, but there is a persisting fact that no one can deny is that space and time are treated very unequally in quantum mechanics. Time enters the Schrodinger equation as a classical variable to parameterize the evolution of the quantum state, while there exists well-defined position operator with its corresponding eigenstate and eigenvalue. This seems to be in contradiction with the spirit of relativity theory, according to which space and time are intimately connected through the Lorentz transformation. Because gravity is nothing but the curvature of spacetime, it is therefore important to have a quantum description of both space and time in order to establish a good quantum gravity theory.

There are some approaches to quantize time in quantum mechanics such as the multiple-time states formalism of Aharonov \textit{et al.} \cite{yakir1,yakir2}, the entangled history formalism of Cotler and Wilczek \cite{cotler}, and the conditional quantum time mechanism of Page and Wootters (PaW) \cite{page}. For a review, see Ref. \cite{review}. Interested readers can confer Ref. \cite{nowakowski} for a possible connection between the multiple-time state and the entangled history formalisms, and Refs. \cite{giovannetti, vedral} for a revised version of the PaW mechanism. See also Refs. \cite{ex1,ex2,ex3} for the prospects of experimental realization of quantum time.

In this paper, we will employ the PaW's proposal of quantum time. According to this model, besides the Hilbert space for the system $\mathcal{H}_S$, there also exists an additional Hilbert space for temporal degrees of freedom $\mathcal{H}_T$, so that the total Hilbert space is $\mathcal{H}_{tot}=\mathcal{H}_T\otimes\mathcal{H}_S$. The states of time $\ket{t}$ live in $\mathcal{H}_T$, the states of the system $\ket{\psi(t)}$ live in $\mathcal{H}_S$, and the total state $\ket{\Psi}\rangle$ lives in $\mathcal{H}_{tot}$. The double ket notation of the total state is just to remind us that it is not the usual state of the system. The total state is specified by the constraint equation $\hat{\mathcal{J}}\ket{\Psi}\rangle=0$, where $\hat{\mathcal{J}}$ is a constraint operator whose explicit form will not be important in our discussion \cite{wheeler}. The main point is that this constraint equation would naturally give rise to the entanglement between the quantum time and the quantum system
\begin{equation}
\ket{\Psi}\rangle=\int dt\ket{t}_T\otimes\ket{\psi(t)}_S.
\end{equation}
The entanglement between time and system is not generated from interaction between them, but is a consequence of the constraint equation $\hat{\mathcal{J}}\ket{\Psi}\rangle=0$ and the fact that it can reproduce the Schrodinger equation by projecting the total state $\ket{\Psi}\rangle$ on the position representation in $\mathcal{H}_T$; that is, $\braket{t|\hat{\mathcal{J}}|\Psi}\rangle=0$ gives the Schrodinger equation. As seen from the ``outside", the global state is static; there is no evolution and there is no time at all. But as seen from the ``inside", there is time because the system evolves and becomes  entangled with time \cite{decoherence}. Thus, the PaW mechanism is also known as ``evolution without evolution". In this quantum time picture,  the more the system evolves, the more it becomes entangled with time.

In this paper, we consider quantum time that is entangled with quantum system due to a Wigner rotation induced by Lorentz transformation. We wish to find the dependence of this time-system entanglement on the rapidity of the Lorentz boost. For this purpose, the qubit clock model discussed in Ref. \cite{qubit} is particularly useful. The total state can be written as
\begin{equation}\label{total state-qubit clock}
\ket{\Psi}\rangle=\frac{1}{\sqrt{2}}\left(\ket{0}_T\ket{\psi_0}_S+\ket{1}_T\ket{\psi_1}_S\right).
\end{equation}
Initially, the system is in the state $\ket{\psi_0}_S$ at the time $\ket{0}_T$. Then, we boost the system in some direction (say, the positive $x$ direction). Effectively, what will happen is that the system's state in $\mathcal{H}_S$ will be rotated by a Wigner rotation, so the system's new state is $\ket{\psi_1}_S$ at the time $\ket{1}_T$. The qubit clock exists in $\mathcal{H}_T$ and it ticks from $\ket{0}_T$ to $\ket{1}_T$ when we make a Lorentz transformation. We note that, under Lorentz transformation, time in the old and new classical reference frames is still classical, but effectively the system's state evolves unitarily in $\mathcal{H}_S$ by the Wigner rotation and we use quantum time to keep track of that evolution. Our work is not yet the full relativistic extension of the PaW mechanism, but is instead a step towards that goal.

\section{Time-System entanglement measures}
 After setting up the stage to study the connection between quantum time and special relativity, we now review the general formulas of time-system entanglement measures. These will include entanglement entropy, mutual information, quadratic entanglement entropy, Renyi entropy, and logarithmic negativity.
 
 The total state in Eq. \ref{total state-qubit clock} is, in general, not a Schmidt decomposition as the states $\ket{\psi_0}_S$ and $\ket{\psi_1}_S$ may not be orthogonal. However, by making a simple transformation, one can write the total state in the form of a Schmidt decomposition \cite{qubit}
\begin{equation}
\ket{\Psi}\rangle=\sqrt{p_+}\ket{+}_T\ket{+}_S+\sqrt{p_-}\ket{-}_T\ket{-}_S,
\end{equation}
where $p_{\pm}=(1\pm|\braket{\psi_0|\psi_1}|)/2$, $\ket{\pm}_T=(\ket{0}_T\pm e^{i\gamma}\ket{1}_T )/\sqrt{2}$, $\ket{\pm}_S=(\ket{\psi_0}_S\pm e^{-i\gamma}\ket{\psi_1}_S)/\sqrt{4p_{\pm}}$, and $e^{i\gamma}=\braket{\psi_0|\psi_1}/|\braket{\psi_0|\psi_1}|$. The entanglement entropy of the time-system entanglement is then 
\begin{equation}\label{ETS equation}
E(T,S)=-\sum_{k=\pm}p_k\log_2p_k.
\end{equation}
As usual in quantum information theory, we use the logarithm in base 2 to measure entropy in units of bits.

The mutual information between two systems is given by $I(A:B)=S(A)+S(B)-S(AB)$ \cite{nelsen}, where $S(A)$ and $S(B)$ are entropies of individual systems, and $S(AB)$ is entropy of the joint system. At the fundamental level, this quantity is particularly relevant to ``extract space from entanglement" \cite{sean1,sean2}. In our case of interest, the global state $\ket{\Psi}\rangle$ is pure. Thus, the mutual information between quantum time and quantum system is given by
\begin{equation}
    I(T,S)=2E(T,S).
\end{equation}

Another useful quantity is quadratic entanglement entropy, which is simply determined from the purity of the density matrix and not its eigenvalues \cite{qubit}
\begin{equation}\label{E2TS}
E_2(T,S)=2(1-\Tr\rho_S^2)=1-|\braket{\psi_0|\psi_1}|^2,
\end{equation}
where $\rho_S$ is the reduced density operator of the system. We included this quantity because purity of the quantum state may be more easily accessible experimentally \cite{tanaka}.

An entanglement measure that recently attracted much attention from both quantum information theorists and quantum field theorists is the Renyi entropy \cite{renyi}. Renyi entropy is given by 
\begin{equation}
    H_n(T,S)=\frac{1}{1-n}\log_2\left(\Tr\rho_S^n\right)=\frac{1}{1-n}\log_2\left(\sum_{k=\pm}p_k^n\right).
\end{equation}
There are a few benchmark values of $n$ that we want to consider. The limit $\lim_{n\rightarrow 0} H_n(T,S)=1$ is sometimes called the Hartley entropy or max-entropy. The limit $\lim_{n\rightarrow 1} H_n(T,S)$ gives the usual entanglement entropy $E(T,S)$ in Eq.\ref{ETS equation}. When $n=2$, $H_2(T,S)$ is sometimes called collision entropy or just Renyi entropy. The limit $\lim_{n\rightarrow\infty}H_n(T,S)=-\log_2(\max[p_k])$ is known as the min-entropy.

The final entanglement measure we would like to discuss is the logarithmic negativity, which is usually considered as one of the most easily computable entanglement measure \cite{vidal}. The logarithmic negativity is given by 
\begin{equation}
    E_\mathcal{N}(T,S)=\log_2(2\mathcal{N}+1)=\log_2(2\sqrt{p_+p_-}+1),
\end{equation}
where $\mathcal{N}$ is the negativity and is computed as the absolute sum of negative eigenvalues of the partial transpose density matrix of the system $\rho^{T_S}$.

All of the above entanglement measures depend on the fidelity. Therefore, once we can calculate the fidelity $|\braket{\psi_0|\psi_1}|$, all entanglement measures can then be determined. The features of fidelity will thus affect the features of entanglement measures, as can be seen clearly in an explicit example presented in the next section.

\section{Example}
As a concrete example, we consider a spin-1/2 particle with the initial state in the momentum representation at time $\ket{0}_T$:
\begin{equation}\label{psi-0}
\psi_0(\textbf{q})=\left(\begin{matrix}
a_1(\textbf{q})\\
a_2(\textbf{q})
\end{matrix}\right),
\end{equation}
and the normalization condition is $\sum_r\int |a_r(\textbf{q})|^2d^3\textbf{q}=1$. Without loss of generality, we can assume that the spin of the particle is aligned in the positive $z$ direction, so that $a_2(\textbf{q})=0$. We further consider a specific case of a Gaussian momentum distribution
\begin{equation}\label{Gaussian}
a_1(\textbf{q})=\frac{e^{-q^2/2w^2}}{\pi^{3/4}w^{3/2}}.
\end{equation}
The parameter $w$ characterizes the spread of momentum. This Gaussian wave function is a minimum uncertainty state and is particularly useful to assess both the quantum and classical limits.

\begin{figure}[h!]
\includegraphics[scale=0.9]{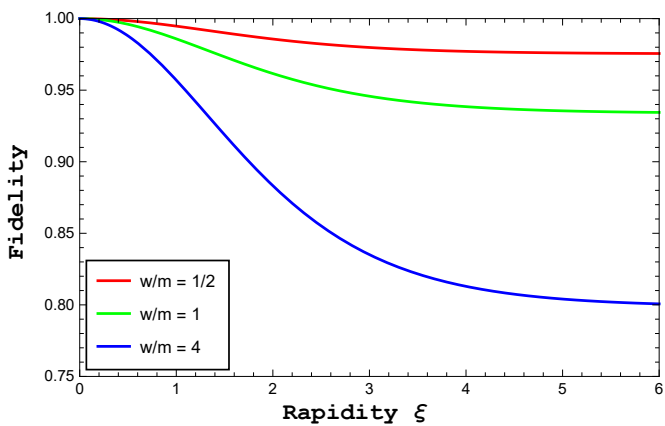}
\caption{Fidelity $|\braket{\psi_0|\psi_1}|$ as a function of rapidity $\xi$ with different values of $w/m$.}
\label{Fidelity}
\end{figure}

Next, we boost this state in the positive $x$ direction. When we do so, the system's state in $\mathcal{H}_S$ will be rotated by a Wigner rotation that acts only on the spin part of the wave function. The unitary operator responsible for this evolution depends on the Lorentz transformation $\Lambda$ and the momentum \textbf{q} of the particle in the system's rest frame \cite{wigner,halpern}:

\begin{widetext}

\begin{equation}\label{Wigner rotation}
U(\Lambda,\textbf{q})=\frac{\cosh\left(\frac{\xi}{2}\right)(E+m)+\sinh\left(\frac{\xi}{2}\right)(\textbf{q}\cdot\hat{\textbf{e}})-i\sinh\left(\frac{\xi}{2}\right)\pmb{\sigma}\cdot(\textbf{q}\times\hat{\textbf{e}})}{\sqrt{(E+m)(E'+m)}},
\end{equation}
where $q^\mu=(E,\textbf{q})$ is the 4-momentum of the particle in the system's rest frame, $p^\mu=\Lambda^\mu_{\ \nu} q^\nu=(E',\textbf{p})$ is the Lorentz boosted momentum, $m$ is the mass of the particle, $\xi$ is rapidity of the boost defined as $\tanh\xi \equiv\beta\equiv v$ (in natural units), $\bm \sigma$ are Pauli spin matrices, and $\hat{\textbf{e}}$ is a unit vector pointing in the direction of the boost ($\hat{\textbf{e}}=\hat{\textbf{x}}$ in our case). We define $q\equiv |\textbf{q}|$ and $p\equiv |\textbf{p}|$. We will use spherical coordinates, so the Wigner rotation is 

\begin{equation}
U(\Lambda,\textbf{q})=K\left(\begin{matrix}
\cosh\left(\frac{\xi}{2}\right)(E+m)+\sinh\left(\frac{\xi}{2}\right)q\sin\theta e^{i\phi} & -\sinh\left(\frac{\xi}{2}\right)q\cos\theta\\
\sinh\left(\frac{\xi}{2}\right)q\cos\theta & \cosh\left(\frac{\xi}{2}\right)(E+m)+\sinh\left(\frac{\xi}{2}\right)q\sin\theta e^{-i\phi}
\end{matrix}\right),
\end{equation}

where $K\equiv \frac{1}{\sqrt{(E+m)(E'+m)}}$ for brevity. We also note that $E'=E\cosh\xi+q\sin\theta\cos\phi\sinh\xi$, so that 
\begin{equation}
K=\frac{1}{\sqrt{(E+m)(E\cosh\xi+q\sin\theta\cos\phi\sinh\xi+m)}}.
\end{equation}
The state of the system at the later time $\ket{1}_T$ after the boost is
\begin{equation}\label{psi-1}
\psi_1(\textbf{p})=\left(\begin{matrix}
b_1(\textbf{p})\\
b_2(\textbf{p})
\end{matrix}\right)=U(\Lambda,\textbf{q})\psi_0(\textbf{q})=K\left(\begin{matrix}
\cosh\left(\frac{\xi}{2}\right)(E+m)+\sinh\left(\frac{\xi}{2}\right)q\sin\theta e^{i\phi}\\
\sinh\left(\frac{\xi}{2}\right)q\cos\theta 
\end{matrix}\right)a_1(\textbf{q}).
\end{equation}

From Eqs. \ref{psi-0} and \ref{psi-1}, the fidelity is

\begin{equation}
|\braket{\psi_0|\psi_1}|=\Bigg|\int K\left[\cosh\left(\frac{\xi}{2}\right)(E+m)+\sinh\left(\frac{\xi}{2}\right)q\sin\theta e^{i\phi}\right]a_1^2(\textbf{q})d^3\textbf{q}\Bigg|.
\end{equation}
\end{widetext}
We substitute the Gaussian wave function in Eq. \ref{Gaussian} and $E=\sqrt{q^2+m^2}$ before performing the integral numerically.
The fidelity is plotted as a function of rapidity $\xi$ in Fig. \ref{Fidelity}. It depends only on the rapidity $\xi$ and the ratio $w/m$.

Fidelity is a measure of overlap between two states in Hilbert space $\mathcal{H}_S$. When the system is not yet boosted $(\xi=0)$, the two states are identical and fidelity is unity. When the system is boosted $(\xi>0)$, the initial and final states become further and further away from each other as the rapidity increases; therefore, fidelity between the two states decreases. But the decrease of fidelity is saturated at some rapidity. Following a similar argument in Ref. \cite{2}, this saturation behavior is explained by the fact that 
$$
\lim_{\xi\rightarrow\infty}K\sinh\left(\frac{\xi}{2}\right)q\cos\theta=\frac{q\cos\theta}{\sqrt{2(E+m)(E+q\sin\theta\cos\phi)}},
$$
and we can maximize this quantity with respect to $\theta$ and $\phi$ to get
$$\left[\lim_{\xi\rightarrow\infty}
K\sinh\left(\frac{\xi}{2}\right)q\cos\theta\right]_{max}= \frac{q/m}{1+\sqrt{1+q^2/m^2}}.
$$ 
This is an increasing function of $q/m$. Roughly speaking, this means that, for a fixed value of $w/m$, there exists a maximum angle that the system's state vector can be rotated and hence the decrease of fidelity is saturated. Also, when we increase the ratio $w/m$, we are also effectively increasing the ratio $q/m$. Thus, the state can be rotated more and more as the maximum angle increases, so that the initial and final states are separated further and further away from each other. This explains why the fidelity decreases stronger if we increase the ratio $w/m$.

\begin{figure}[h!]
\includegraphics[scale=0.9]{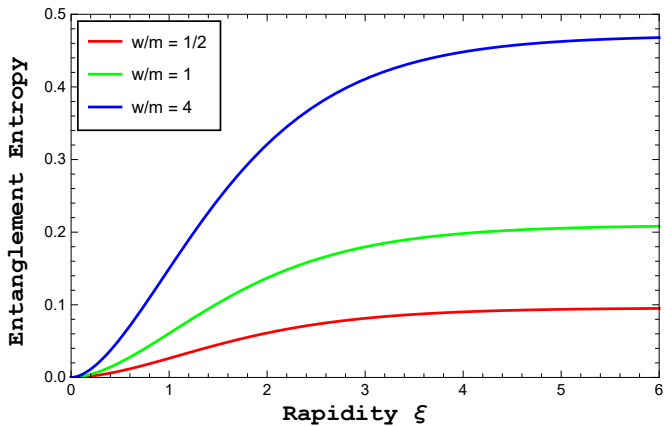}
\caption{Time-System entanglement entropy $E(T,S)$ as a function of rapidity $\xi$ with different values of $w/m$.}
\label{ETS}
\end{figure}

The time-system entanglement entropy $E(T,S)$ can then be calculated from the fidelity using Eq. \ref{ETS equation} and is plotted as a function of rapidity $\xi$ in Fig. \ref{ETS}. We see that the time-system entanglement entropy increases when rapidity increases. That is because the more the system evolves, the more it becomes entangled with time. When the rapidity increases, the initial and final states are more distant from each other (which means smaller fidelity), so the system must travel through a larger distance in Hilbert space and hence it becomes entangled more strongly with time. The saturation of $E(T,S)$ is explained similarly to the case of fidelity. Specifically, for a fixed value of $w/m$, there exists a maximum distance between the initial and final states, so that the time-system entanglement entropy is saturated at some rapidity. If we increase the ratio $w/m$, this maximum distance increases and hence the time-system entanglement entropy increases stronger. Also note that in the classical limit when $w/m\ll 1$ (i.e. the particle is more massive and has more definite momentum), the entanglement entropy is very small and eventually disappears when $w/m\rightarrow 0$. That is because entanglement entropy is a signature of the quantumness of the system and is not present classically.

\begin{figure}[h!]
    \centering
    \includegraphics[scale=0.9]{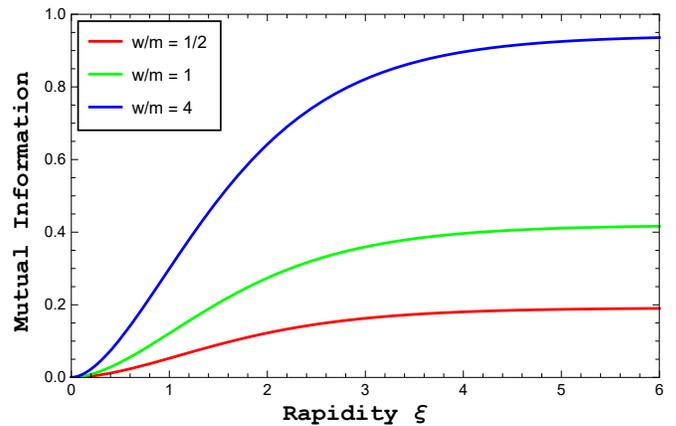}
    \caption{Mutual information of time-system entanglement $I(T,S)$ as a function of rapidity $\xi$ for different values of $w/m$.}
    \label{fig:MutualInformation}
\end{figure}

\begin{figure}[h!]
\includegraphics[scale=0.9]{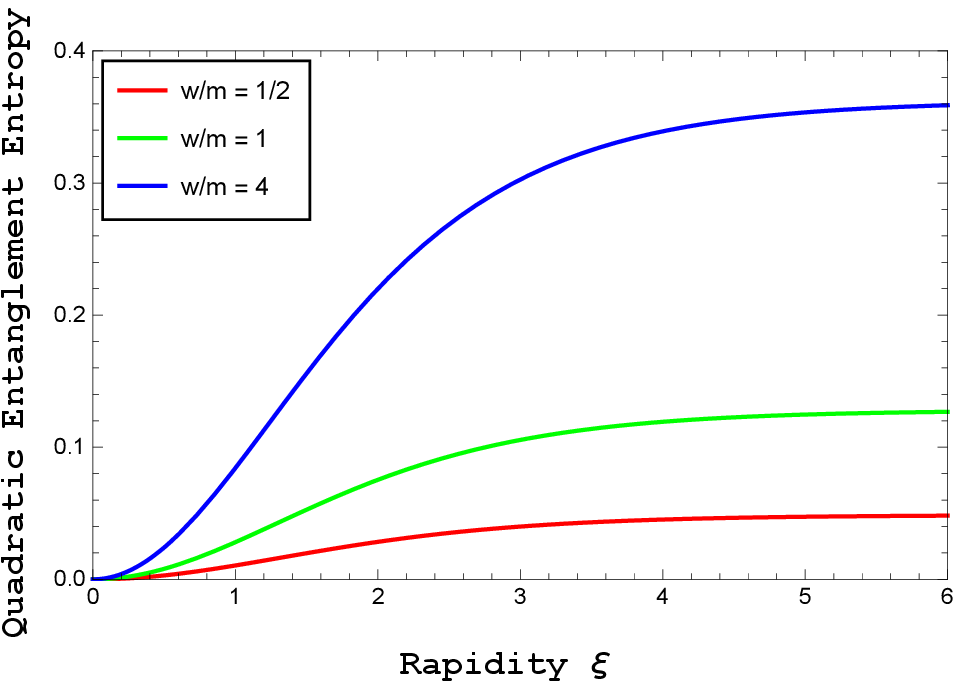}
\caption{Time-System quadratic entanglement entropy $E_2(T,S)$ as a function of rapidity $\xi$ with different values of $w/m$.}
\label{E2TS-figure}
\end{figure}

\begin{figure}[h!]
    \centering
    \includegraphics[scale=0.9]{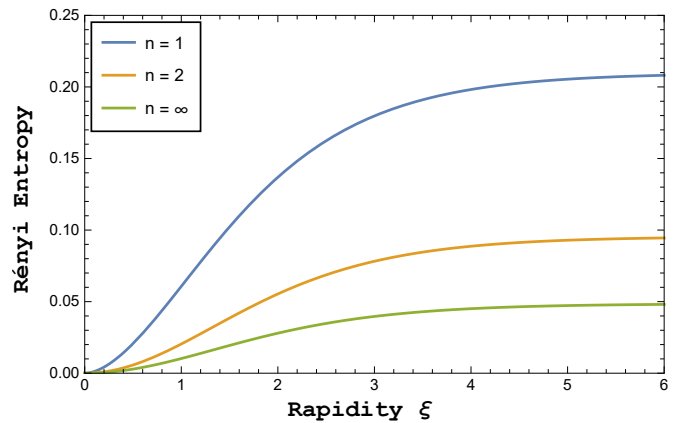}
    \caption{Renyi entropy of time-system entanglement $H_n(T,S)$ as a function of rapidity $\xi$ for different values of $n$ and with $w/m=1$. The case $n=1$ is exactly $E(T,S)$ shown in Fig. \ref{ETS}.}
    \label{fig:renyi entropy}
\end{figure}

\begin{figure}[h!]
    \centering
    \includegraphics[scale=0.9]{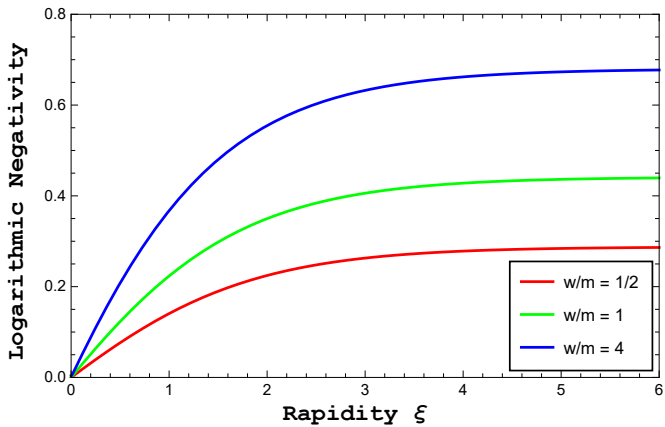}
    \caption{Logarithmic negativity of time-system entanglement $E_\mathcal{N}(T,S)$ as a function of rapidity $\xi$ for different values of $w/m$.}
    \label{fig:logarithmic negativity}
\end{figure}

The mutual information, quadratic entanglement entropy, Renyi entropy, and logarithmic negativity of time-system entanglement are  shown in Figs. \ref{fig:MutualInformation}, \ref{E2TS-figure}, \ref{fig:renyi entropy}, \ref{fig:logarithmic negativity} respectively. We see that these entanglement measures share similar features to the standard entanglement entropy (Fig. \ref{ETS}) such as the increasing and saturation behaviours, though the scales between them, of course, differ from each other.

\begin{figure}[h!]
\includegraphics[scale=0.9]{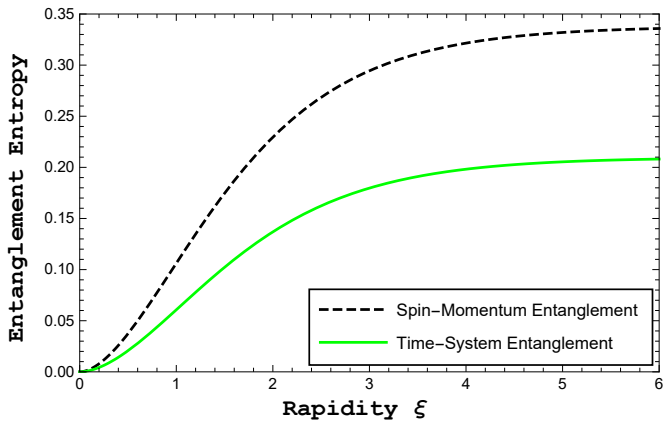}
\caption{Comparison between time-system entanglement entropy and spin-momentum entanglement entropy. As an illustration, we chose $w/m=1$ for both quantities.}
\label{Comparison}
\end{figure}

It is also known that when the initial, unentangled state of a particle is Lorentz boosted, there will be entanglement between the spin and  momentum degrees of freedom of that state \cite{1}. It is therefore interesting to compare time-system entanglement with spin-momentum entanglement. The spin-momentum entanglement entropy is \cite{1}
\begin{equation}
S=-\sum_k\lambda_k\log_2\lambda_k,
\end{equation}
where $\lambda_k=(1\pm|\textbf{n}|)/2$ and $\textbf{n}$ is the Bloch vector whose components are $n_x=n_y=0$ and 
\begin{equation}
n_z=\int\left(|b_1(\textbf{p})|^2-|b_2(\textbf{p})|^2\right)d^3\textbf{p}.
\end{equation}
The functions $b_1(\textbf{p})$ and $b_2(\textbf{p})$ are given in Eq. \ref{psi-1}. Note that here we are doing the integral over the momentum variable $\textbf{p}$ in the boosted frame, so that formally we need to rescale the factor $K$ as follows: $K\rightarrow K(E/E')^{1/2}$. This is just to ensure that we can convert a Lorentz invariant integration measure $d^3\textbf{p}/2E'$ to $d^3\textbf{q}/2E$ before performing the integral over \textbf{q} numerically. The comparison of entanglement entropy between spin-momentum entanglement and time-system entanglement is plotted in Fig. \ref{Comparison}. As an illustration, we chose $w/m=1$ for both quantities. We see that the time-system entanglement entropy and the spin-momentum entanglement entropy have a similar saturation pattern, and that the former is smaller than the latter. Note that the spin-momentum entanglement happens \textit{within} the system, while the system as a whole becomes entangled with time as it evolves.

\section{Summary}
In summary, we considered a qubit clock that is entangled with a quantum system due to the Wigner rotation induced by Lorentz transformation. Our purpose is to find the dependence of this time-system entanglement on the rapidity of the Lorentz boost. Specifically, we considered the case of a spin-1/2 particle with Gaussian momentum distribution. We obtained the result that the time-system entanglement entropy will increase when rapidity increases. Intuitively, this can be explained by the fact that the two initial and final states are more distant from each other for larger rapidity, so that the system must travel through a larger distance in Hilbert space and hence it becomes more entangled with time. This feature reflects the spirit of the Page-Wootters quantum time proposal that the evolution of the quantum system is described by the entanglement between it and quantum time. We also compared this novel time-system entanglement entropy with the well-known spin-momentum entanglement entropy and find that the former is smaller than the latter.

In addition, we computed some other interesting time-system entanglement measures such as mutual information, quadratic entanglement entropy, Renyi entropy, and logarithmic negativity. Due to the fact that they all depend on the fidelity, these entanglement measures share similar features to the standard entanglement entropy such as the increasing and saturation behaviours.

A natural possible extension of our work would be computing time-system entanglement measures for the case of two-particle or many-particle systems when they are boosted. Another direction is to explore further the implications of quantum time for quantum field theory \cite{QFT1,QFT2,QFT3}, quantum gravity \cite{carlo3,zeh,salecker} and cosmology \cite{kiefer}. Finding the quantum origin of the arrow of time is also an interesting and ongoing research \cite{lorenzo,david,justin,john,zeh2}. Recently, there were also some works on time-system entanglement when the system itself contains entangled subsystems \cite{loc23}, or the system interacts with the environment \cite{sebas}. These are compelling topics that could be relevant for exploring the insights of quantum time into the information paradox of black hole \cite{hawking} or cosmology \cite{cosmo}. We hope to address these ideas in future works.


\begin{thebibliography}{99}
\bibitem{wheeler} It is also interesting to note that this equation mimics the timeless Wheeler-DeWitt equation of quantum gravity \cite{dewitt}. See also  \cite{carlo1}.
\bibitem{decoherence} In a sense, this is quite similar to the spirit of decoherence (for example, see \cite{joos}): as seen from the ``outside", the Universe is just an abstract quantum state and there is nothing like stars, galaxies, or people; but as seen from the ``inside", the stuff constantly interacts with each other, so that decoherence happens and the appearance of the classical world is emerged naturally. Decoherence is actually an example of weak emergence, which means that the macroscopic laws are expected to be derivable from the more fundamental underlying microscopic laws.
\end{thebibliography}
\end{document}